\documentclass[prl,twocolumn,showpacs,preprintnumbers]{revtex4}
\usepackage{rotate}
\usepackage{epsf}

\newcommand{\insertfig}[2]{\mbox{\epsfxsize=#1cm \epsfbox{#2.eps}}}
\newcommand{\bit}[1]{\mbox{\boldmath$#1$}}

\begin{document}

\preprint{DOE/ER/40762-271 \ UM-PP\#03-031}

\title{Perturbative QCD analysis of the nucleon's Pauli form factor $F_2(Q^2)$}

\author{A.V. Belitsky, Xiangdong Ji, and Feng Yuan}

\affiliation{Department of Physics, University of Maryland,
             MD 20742-4111, College Park, USA}

\begin{abstract}
We perform a perturbative QCD analysis of the nucleon's Pauli
form factor $F_2(Q^2)$ in the asymptotically large $Q^2$ limit.
We find that the leading contribution to $F_2(Q^2)$ has a $1/Q^6$
power behavior, consistent with the well-known result in the literature.
Its coefficient depends on the leading- and subleading-twist light-cone
wave functions of the nucleon, the latter describing the quarks with one unit of
orbital angular momentum. We also derive at the logarithmic
accurary the asymptotic scaling $F_2(Q^2)/F_1(Q^2) \sim (\log^2
Q^2/\Lambda^2)/Q^2$ which describes recent Jefferson Lab data
well.
\end{abstract}


\maketitle

The electromagnetic form factors are fundamental observables of the nucleon
containing important information about its internal nonperturbative structure.
Since the first measurement in the mid 1950s, experimental studies of these
observables have become an active frontier in nuclear and particle physics. More
recently, with the development of novel experimental techniques, the magnetic
form factor of the proton, and electric and magnetic form factors of the neutron
have been measured with unprecedented precision at Jefferson Lab and other
facilities around the world \cite{Gao03}. In particular, the Pauli form factor
of the proton $F_2$ extracted from the recoil polarizations at Jefferson Lab
reveals a significant difference from previous data and theoretical expectations
\cite{Jon00,Gay02}. The data shows that the ratio of Dirac to Pauli form factor
$Q^2 F_2(Q^2)/F_1(Q^2)$ continues climbing at the largest $Q^2$'s measured and
seems to scale as $\sqrt{Q^2}$. The result has sparked myriad speculations
on its implication about the underlying microscopic structure
of the proton \cite{BraLenMahSte01,Mil02,RalJai02,MaQinSch02,Bro02}.

\begin{figure}[htb]
\begin{center}
\mbox{
\begin{picture}(0,70)(85,0)
\put(-5,0){\insertfig{6}{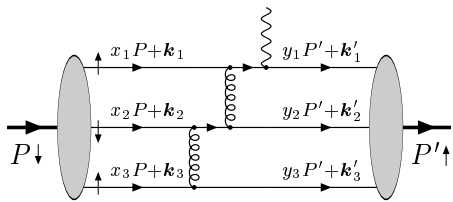}}
\end{picture}
}
\end{center}
\vspace*{-0.3cm}
\caption{\label{Typical} A leading QCD diagram contributing to the nucleon's
form factors.}
\end{figure}

A simple dimensional counting rule was devised in Refs.\ \cite{BroFar73,MatMurTav73}
to determine the dominant power contribution to hadronic form factors at
large momentum transfer. For the hadron helicity-conserving form factor
$F_1(Q^2)$, it predicts a dominant scaling behavior $1/Q^4$. The power
counting can be justified by QCD factorization theorems which separate
short-distance quark-gluon interactions from soft hadron wave functions
\cite{LepBro80,EfrRad80,CheZhi77,DunMue79,AznEsaAtsTer80,AvdKorChe81,CarGro87,Ste99},
see Fig.\ \ref{Typical}. Recently, a more sophisticated formalism has been
developed to include higher-order perturbative QCD (PQCD) resummation, or
Sudakov form factor, which treats contributions from the small-$x$ partons
more appropriately \cite{DunMue79,Rad84,IsgLle89,LiSte92,Li93}. The procedure extends
the applicability of leading-order PQCD predictions to moderate $Q^2$, but
does not change the dominant power-law behavior significantly
\cite{BolJacKroBerSte95,Li93,Ste99}.

The literature on PQCD studies of $F_2(Q^2)$, however, is much meager. Since
$F_2 (Q^2)$ is related to the hadron helicity-flip amplitude, its power behavior
is suppressed compared to $F_1(Q^2)\sim 1/Q^4$, and a natural expectation is
then $1/Q^6$. This is confirmed by calculations where the quark masses are
the mechanism for the helicity flip \cite{BroFar73,JiSil86}. Since the up and down
current quark masses are negligible, the dominant mechanism for the spin flip
in QCD come from the quark orbital angular momentum and the polarization
of an extra gluon which must be included to maintain gauge invariance \cite{BroDre80}.
A generalized power counting including the parton orbital angular momentum
validates the expected scaling law for $F_2(Q^2)$ \cite{JiMaYua03}. An actual
PQCD calculation for $F_2(Q^2)$ requires the
extension of the usual formalism to include the quark orbital angular
momentum, a technology has not yet been systematically developed in
the literature. The recent development in classification of the
light-cone wave functions \cite{BurJiYua02,JiMaYua02} provides the
necessary ingredient to perform these calculations.

In this paper we report on a first PQCD calculation of $F_2(Q^2)$ at
asymptotically large momentum transfer $Q^2$. Modulo logarithms, we confirm
that the leading contribution to $F_2(Q^2)$ goes like $1/Q^6$. The coefficient
of the leading-power term depends on leading order (twist-three) and
next-to-leading order (twist-four) light-cone wave functions. The latter is
the probability amplitude for one of the quarks to carry one unit of orbital
angular momentum. We compute the leading-order perturbative kernel, and
estimate the coefficient using models for the light-cone wave functions.
We comment on the role of the Sudakov form factor in regulating possible
end-point singularities in the phase space integrals. We also derive with
the logarithmic accuracy the asymptotic scaling $F_2(Q^2)/F_1(Q^2)\sim
(\log^2Q^2/\Lambda^2)/Q^2$ which we find to describe the data
unexpectedly well.

To start with, we choose a coordinate system for the process $P+\gamma^*
\rightarrow P'$ such that $P$ and $P'$ form conjugate light-cone vectors.
We let the light-cone components $P^+$ and ${P'}^-$ to be large in the
asymptotic limit, $Q^2 = 2 P^+ {P'}^-$. The $F_2(Q^2)$ form factor can be
extracted from the helicity-flip matrix element
\begin{equation}
\langle P'_\uparrow | J^\mu |P_\downarrow \rangle
=
F_2 (Q^2)
\bar u_\uparrow (P') \frac{i\sigma^{\mu\alpha} q_\alpha}{2M} u_\downarrow (P)
\, ,
\end{equation}
where $J^\mu$ is the quark electromagnetic current and $u (P)$ is an on-shell
spinor of the nucleon with momentum $P$. In the above coordinate system, the
spin of the nucleon actually remains the same in the initial and final states.

A perturbative analysis of the above matrix element is done by computing
leading-order Feynman diagrams with two gluon exchanges, for a typical graph
see Fig.\ \ref{Typical}, in which the initial (final) nucleon contains three
quarks with longitudinal momenta $x_i P$ ($y_i P'$) and transverse momenta
$\bit{k}_i$ ($\bit{k}'_i$) of order $\Lambda_{\rm QCD}$. Two hard-gluon exchanges
ensure that the three quarks in the final state propagate collinearly after the
injection of a large momentum transfer $q^\mu$. Since up and down quarks are
light, quark helicity is approximately conserved during the hard scattering.
Therefore, to produce a nucleon-spin flip, the quark orbital angular momentum in
the initial and final states must differ by one unit. The leading contribution
to $F_2(Q^2)$ comes from the configurations in which the quarks in the initial
or the final state carry zero unit, and those in the other state carry one unit,
of orbital angular momentum. For definiteness, let us assume that the final
state quarks are symmetric in azimuth.

To isolate the leading contribution, we expand the hard part of the diagram in
$\bit{k}^2_i/Q^2$ in the limit of large $Q^2$ and $x_i \ne 0$ (a procedure
dubbed the collinear expansion). We will comment on the $x_i\rightarrow 0$ case
later. Since the final state nucleon has no orbital angular momentum, we throw
away all the subleading terms in $\bit{k}_i'$. On the other hand, the quarks in
the initial state nucleon have one unit of orbital angular momentum, and hence
the hard part must have linear terms in the quark transverse momenta. Therefore,
the leading hard part, we are interested in, has a structure $\bit{k}_i T(x_1,x_2,x_3,
y_1,y_2,y_3; Q^2)$.

Let us consider the simplification of the nucleon wave functions after the
collinear expansion. For definiteness, we consider the proton form factor. The
light-cone wave function for the final state is
\begin{eqnarray}
&&\!\!\!\!\!\!{| P_\uparrow \rangle}_{1/2}
=
\frac{1}{12}
\int \frac{[d x] [d^2 \bit{k}]}{\sqrt{x_1 x_2 x_3}} \,
\psi_1 (\kappa_1, \kappa_2, \kappa_3)
\\
&&\ \times
\varepsilon^{abc} \,
u^\dagger_{a \uparrow} (\kappa_1)
\left\{
u^\dagger_{b \downarrow} (\kappa_2)
d^\dagger_{c \uparrow} (\kappa_3)
-
d^\dagger_{b \downarrow} (\kappa_2)
u^\dagger_{c \uparrow} (\kappa_3)
\right\}
| 0 \rangle \, ,
\label{w1}
\nonumber
\end{eqnarray}
where the argument $\kappa_i$ is a shorthand notation for $(x_i, \bit{k}_i)$,
and the integration measures for the quark momenta are
\begin{eqnarray*}
&&{[d x]}
\equiv
d x_1 \, d x_2 \, d x_3 \,
\delta \left( x_1 + x_2 + x_3 - 1 \right)
\, , \\
&&{[d^2 \bit{k}]}
\equiv
\,
d^2 \bit{k}_1 \, d^2 \bit{k}_2 \, d^2 \bit{k}_3 \,
\delta^{(2)} \left( \bit{k}_1 + \bit{k}_2 + \bit{k}_3 \right)
\, .
\end{eqnarray*}
If the hard part has no dependence on $\bit{k}'_i$, we can ignore the
transverse momentum dependence in the quark creation operators and define
the twist-three amplitude
\begin{eqnarray*}
{\mit\Phi}_3 (x_1, x_2, x_3)
=
2 \int [d^2 \bit{k}] \,
\psi_1 (\kappa_1, \kappa_2, \kappa_3)
\, .
\end{eqnarray*}
Then the final nucleon state can be simplified to,
\begin{eqnarray}
&&\!\!\!\!\!\!{| P_\uparrow \rangle}_{1/2}
=
\frac{1}{24}
\int \frac{[d x]}{\sqrt{x_1 x_2 x_3}} \,
{\mit\Phi}_3 (x_1, x_2, x_3)
\\
&&\ \times
\varepsilon^{abc} \,
u^\dagger_{a \uparrow} (x_1)
\left\{
u^\dagger_{b \downarrow} (x_2)
d^\dagger_{c \uparrow} (x_3)
-
d^\dagger_{b \downarrow} (x_2)
u^\dagger_{c \uparrow} (x_3)
\right\}
| 0 \rangle \, .
\nonumber
\end{eqnarray}
In the first quantization formalism, the second line can be replaced by the
standard $SU (6)$ wave function.

Since the quarks in the initial nucleon state must have a total helicity
$1/2$, only the following wave function component is relevant
\cite{JiMaYua02,BurJiYua02}
\begin{eqnarray}
&&\!\!\!\!\!\!\!\!\!\!{| P_\downarrow \rangle}_{1/2}
=
\frac{1}{12}
\int \frac{[d x] [d^2 \bit{k}]}{\sqrt{x_1 x_2 x_3}} \,
\left\{ \bar k_{1 \perp} \psi_3 + \bar k_{2 \perp} \psi_4 \right\}
(\kappa_1, \kappa_2, \kappa_3)
\nonumber\\
&&\!\!\!\!\!\!\!\!\times
\varepsilon^{abc} \,
u_{a \uparrow}^\dagger (\kappa_2)
\left\{
u_{b \downarrow}^\dagger (\kappa_1)
d_{c \uparrow}^\dagger (\kappa_3)
-
d_{b \downarrow}^\dagger (\kappa_1)
u_{c \uparrow}^\dagger (\kappa_3)
\right\}
| 0 \rangle ,
\end{eqnarray}
where we use the notation $\bar k_\perp \equiv k^x - i k^y$.

The spin-isospin structure of this wave function is exactly the same as that
in Eq.\ (\ref{w1}). After the collinear expansion, the above wave function
will be convoluted with transverse momenta of quarks. We define the twist-four
amplitudes via
\begin{eqnarray*}
{\mit\Phi}_4 (x_2, x_1, x_3)
\!\!\!&=&\!\!\!
2 \!\int \frac{[d^2 \bit{k}]}{M x_3} \,
\bit{k}_3 \cdot\!
\left\{
\bit{k}_1 \psi_3
+
\bit{k}_2 \psi_4
\right\}
(\kappa_1, \kappa_2, \kappa_3)
, \\
{\mit\Psi}_4 (x_1, x_2, x_3)
\!\!\!&=&\!\!\!
2 \!\int \frac{[d^2 \bit{k}]}{M x_2} \,
\bit{k}_2 \cdot\!
\left\{
\bit{k}_1 \psi_3
+
\bit{k}_2 \psi_4
\right\}
(\kappa_1, \kappa_2, \kappa_3)
.
\end{eqnarray*}
Apart from a gluon-potential dependent term, the above light-cone amplitudes
are the same as those defined in Ref.\ \cite{BraFriMahSte00}. In the following,
we include the former so that the result is gauge invariant. To the order that
we are working, the gluon potential terms arise from perturbative diagrams
with explicit gluons attached to the nucleon blobs. These contributions are
generally considered as dynamically suppressed \cite{Pol80}.

We introduce effective nucleon wave functions for the initial state after
integrating over the transverse momenta weighted with a momentum factor from
the hard part,
\begin{eqnarray}
&&\!\!\!\!\!\!{| P_\downarrow \rangle}_{1/2} [\bit{k}_1]
= \frac{M}{24}
\int \frac{[d x]}{\sqrt{x_1 x_2 x_3}} \, x_1 {\mit\Psi}_4 (x_2, x_1, x_3)
\\
&&\times\epsilon^{abc}
u^{\dagger}_{a \uparrow}(x_1)
\left\{
u^{\dagger}_{b \downarrow} (x_2) d^{\dagger}_{c \uparrow} (x_3)
-
d^{\dagger}_{b \downarrow} (x_2) u^{\dagger}_{c \uparrow} (x_3)
\right\}
|0\rangle
\, , \nonumber \\
&&\!\!\!\!\!\!{| P_\downarrow \rangle}_{1/2} [\bit{k}_3]
= \frac{M}{24}
\int \frac{[d x]}{\sqrt{x_1 x_2 x_3}} \, x_3 {\mit\Phi}_4 (x_1, x_2, x_3)
\\
&&\times
\epsilon^{abc}
u^{\dagger}_{a\uparrow} (x_1)
\left\{
u^{\dagger}_{b\downarrow} (x_2) d^{\dagger}_{c\uparrow} (x_3)
-
d^{\dagger}_{b\downarrow} (x_2) u^{\dagger}_{c\uparrow} (x_3)
\right\}
|0\rangle
\, , \nonumber
\end{eqnarray}
where the argument on the left-hand side indicates the momentum being averaged.
Averaging over $\bit{k}_2$ does not yield independent information.

Using the twist-3 and -4 amplitudes, we can write the $F_2(Q^2)$ form factor in the
following factorized form,
\begin{eqnarray}
F_2(Q^2)
\!&=&\!\!
\int [dx] [dy] \Big\{
x_3 {\mit\Phi}_4 (x_1, x_2, x_3) T_{\mit\Phi} (\{x\}, \{y\})
\\
&+&\!\!
x_1 {\mit\Psi}_4 (x_2, x_1, x_3) T_{\mit\Psi} (\{x\}, \{y\})
\Big\}
{\mit\Phi}_3 (y_1, y_2, y_3)
\, , \nonumber
\end{eqnarray}
where $\{x\}=(x_1,x_2,x_3)$. Let us see how to extract the hard part,
$T$, in Fig.\ \ref{Typical}.

We use the nomenclature and strategy of Ref.\ \cite{LepBro80} by calling the
top quark line 1, the middle 2, etc., without committing them to a specific
flavor. Given the spin-isospin structure of the initial and final state
nucleon wave functions, we assume that the first and third quarks have spin
up and the second one spin down. Call the amplitudes $T_i$ when the
electromagnetic current acts on particle $i$. Clearly because of symmetry,
$T_3$ can be obtained from $T_1$ by a suitable exchange of variables (see
below). Using the $SU (6)$ wave function and the quark charge weighting,
we find the flavor structure of the hard part for the proton,
\begin{eqnarray}
T^p (\{x\}, \{y\})
\!\!&=&\!\!
\frac{2e_u}{3}T_1
+ \frac{e_u + e_d}{3} (T_2+ T_3)
\nonumber\\
&+&\!\!
\frac{e_u}{3}(T_1'+T_3') + \frac{e_d}{3} T_2'
\, .
\end{eqnarray}
where we have omitted the argument of $T_i$ on the right-hand side, and
$T'_i$ has $y_1$ and $y_3$ interchanged. For the neutron, $u\leftrightarrow d$.

According to the above convention, we find the contribution to $T_1$ from
Fig.\ \ref{Typical},
\begin{eqnarray}
T_{1{\mit\Psi}}^{\rm fig.1} (\{x\}, \{y\})
&=&
- \frac{M^2 C_B^2}{Q^6}
(4\pi\alpha_s)^2
\frac{1}{{\bar x}_1^2 x_3 {\bar y}_1 y_3^2}
\, , \nonumber \\
T_{1{\mit\Phi}}^{\rm fig.1} (\{x\}, \{y\})
&=&
- \frac{M^2C_B^2}{Q^6}
(4\pi\alpha_s)^2
\frac{1}{ {\bar x}_1 x_3^2 {\bar y}_1y_3^2}
\, ,
\end{eqnarray}
where here and everywhere $\bar x \equiv 1 - x$. $C_B = 2/3$ is a color factor
and the contribution from the twist-4 wave functions of the final-state nucleon
has also been included. The corresponding contribution to $T_3$ can be obtained
by interchanging the labels, $T_{3{\mit\Psi}}^{\rm fig.1} =  T_{1{\mit\Phi}}^{\rm fig.1}
(1\leftrightarrow 3)$, $T_{3{\mit\Phi}}^{\rm fig.1} =  T_{1{\mit\Psi}}^{\rm fig.1}
(1 \leftrightarrow 3)$. This in fact is a general feature.

\begin{figure}[htb]
\begin{center}
\mbox{
\begin{picture}(0,135)(90,0)
\put(-5,0){\insertfig{7}{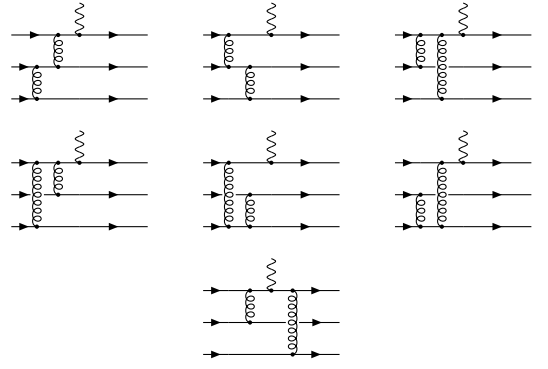}}
\end{picture}
}
\end{center}
\vspace*{-0.3cm}
\caption{\label{set} Perturbative diagrams contributing to the hard part
of $F_2$. Mirror symmetric graphs have to added.}
\end{figure}

To find a complete expression for the hard part, we must calculate perturbative
diagrams displayed in Fig.\ \ref{set}. A straightforward evaluation yields,
\begin{eqnarray}
T_{1 {\mit\Psi}}(\{x\}, \{y\})
&=&\!
\frac{M^2C_B^2 }{Q^6}
(4\pi\alpha_s)^2 \, x_3(G_{11}-G_{12})
\, , \\
T_{1 {\mit\Phi}}(\{x\}, \{y\})
&=&\!
\frac{M^2C_B^2 }{ Q^6}
(4\pi\alpha_s)^2 \, [ - x_1 G_{11} - {\bar x}_1 G_{12} ]
\, , \nonumber\\
T_{2 {\mit\Psi}} (\{x\}, \{y\})
&=&\!
\frac{M^2C_B^2 }{Q^6}
(4\pi\alpha_s)^2
\nonumber\\
&\times&\! [ x_3(G_{22} - \widetilde G_{21} - \widetilde G_{22}) - {\bar x}_3 G_{21} ]
\, , \nonumber
\end{eqnarray}
and $T_{2{\mit\Phi}} = T_{2{\mit\Psi}} (1 \leftrightarrow 3)$, $T_{3{\mit\Psi}}
= T_{1{\mit\Phi}} (1 \leftrightarrow 3)$, and $T_{3{\mit\Phi}} = T_{1{\mit\Psi}}
(1 \leftrightarrow 3)$. The functions $G_{ij}$ are defined as
\begin{eqnarray}
G_{11}
&=&
\frac{1}{x_1 x_2 x_3^2 y_2 y_3^2 {\bar y}_3}
\, , \nonumber\\
G_{12}
&=&
\frac{1}{x_3^2 {\bar x}_1^2 y_3 {\bar y}_1^2}
+
\frac{1}{x_2 x_3 {\bar x}_1^2 y_2 {\bar y}_1^2}
\nonumber \\
&&
+ \frac{1}{ {\bar x}_1^2 x_3^2 {\bar y}_1 y_3^2}
- \frac{1}{x_2 x_3^2 {\bar x}_1 y_2 y_3 {\bar y}_3}
\, , \nonumber\\
G_{21}
&=&
\frac{1}{x_1^2 x_3 {\bar x}_3 y_1^2 y_3 {\bar y}_1}
\, , \nonumber\\
G_{22}
&=&
\frac{1}{x_1 x_3^2 {\bar x}_2 y_3^2 {\bar y}_2}
+
\frac{1}{x_1^2 x_3^2 y_1 y_3 {\bar y}_1}
\, ,
\end{eqnarray}
and $\widetilde G_{21} = G_{21} (1 \leftrightarrow 3)$, $\widetilde G_{22} =
G_{22}(1 \leftrightarrow 3)$.

To determine the normalization for $F_2(Q^2)$, we need to know the light-cone
distribution amplitudes ${\mit\Phi}_3$, ${\mit\Phi}_4$ and ${\mit\Psi}_4$,
which can only be obtained by solving QCD nonperturbatively. However, the
scale evolution of these amplitudes selects at the asymptotically
large $Q^2$ the leading component with a fixed small-$x_i$ behavior. For
example, the asymptotic form of ${\mit\Phi}_3$ is $x_1 x_2 x_3$ \cite{LepBro80},
whereas that of ${\mit\Phi}_4$ is $x_1 x_2$. In Ref.\ \cite{BraFriMahSte00}
a set of phenomenological amplitudes satisfying these asymptotic constraints
have been proposed on a basis of conformal expansion.

With the above wave functions, the integrals over momentum fractions $x_i$
and $y_i$ have logarithmic singularities, indicating that
the factorization breaks down when one of the quarks in the wave function
becomes soft \cite{Rad84,IsgLle89}. It has been suggested that the higher-order
PQCD resummation, or the Sudakov form factor, suppresses the contribution at
small-$x$ and provides an effective cut-off for the integrals at $x \sim
\Lambda^2 / Q^2$, where $\Lambda$ is a soft scale related to the size of the
nucleon \cite{Li93,LiSte92,BolJacKroBerSte95,MusRad97,DesSac01}. The outcome
is that the $x_i$ integrations contribute an extra $Q^2$-dependent factor
$\ln^2 Q^2/\Lambda^2$, compared to $F_1(Q^2)$. Physically, the end-point
divergencies indicate that quarks with different rapidity contribute equally
to the hard scattering. Since the contribution from quarks with very large
rapidity (small-$x$) is suppressed by the Sudakov form factor, this
$Q^2$-dependence reflects simply the {\sl kinematic broadening of the quark
(and gluon) rapidity range with increasing nucleon momentum}.

For an estimate, we use asymptotic wave functions \cite{BraFriMahSte00},
\begin{eqnarray}
&&{\mit\Phi}_3
=
120 \, x_1 x_2 x_3 f_N
\, , \
{\mit\Phi}_4
=
12 \, x_1 x_2 (f_N + \lambda_1)
\, , \nonumber\\
&&{\mit\Psi}_4
=
12 \, x_1 x_3 (f_N - \lambda_1) \, ,
\end{eqnarray}
with $f_N = 5.3 \cdot 10^{-3} \ {\rm GeV}^2$ and $\lambda_1 = - 2.7 \cdot 10^{-2}
\ {\rm GeV}^2$. With a choice of $\Lambda = (0.3 \ {\rm GeV})^2$, $Q^6 F_2^p (Q^2)$
is roughly $0.6 \ {\rm GeV}^6$ for $Q^2 = (5-20) \ {\rm GeV}^2$, about $1/3$ of
the Jefferson Lab data at $Q^2 = 5 \ {\rm GeV}^2$ \cite{Gay02}. Of course, to get
a more realistic PQCD prediction in this regime, one must have the quark
distribution amplitudes appropriate at this scale. However, from the comparison
between the data and PQCD predictions for $F_1(Q^2)$ \cite{Bro02}, we believe that asymptotic
PQCD is unlikely to be the dominant contribution to $F_2(Q^2)$ at $Q^2 = 3 - 5 \,
{\rm GeV}^2$: one must take into account higher-order corrections and higher-twist
effects.

\begin{figure}[htb]
\begin{center}
\mbox{
\begin{picture}(0,155)(120,0)
\put(-5,0){\insertfig{8}{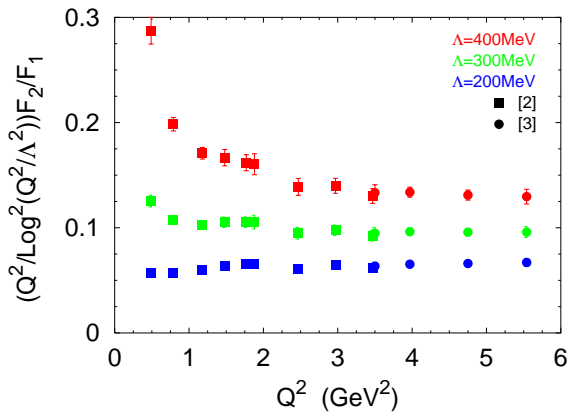}}
\end{picture}
}
\end{center}
\vspace*{-0.3cm}
\caption{\label{ScalingF2toF1} JLab data plotted in terms of the
leading PQCD scaling. The low, middle, and upper data points
correspond to $\Lambda=200,300,400$, respectively. }
\end{figure}

Coming back to the scaling behavior of the ratio $F_2(Q^2)/F_1(Q^2)$ for which
the Jefferson Lab data has stimulated much discussion in the literature. PQCD
predicts the power-law scaling $1/Q^2$. With the new result for $F_2(Q^2)$, we can
determine its scaling up to logarithmic accuracy. The strong coupling constant
in the ratio simply cancels. The wave function evolution yields a factor of
$\alpha_s^{32/(9 \beta)} (Q^2)$ for $F_1 (Q^2)$ and $\alpha_s^{8/(3 \beta)}
(Q^2)$ for $F_2 (Q^2)$ from the leading non-vanishing contribution, where
$\beta = 11 - 2 n_f/3$. Thus PQCD predicts that
$( Q^2/\ln^{2 + 8/(9 \beta)} Q^2/\Lambda^2 )(F_2(Q^2)/F_1(Q^2))$ scales as
a constant at large $Q^2$, $8/(9 \beta) \ll 1$. Surprisingly, the Jefferson
Lab data plotted this way (ignoring the small $8/(9\beta)$)
exhibits little $Q^2$ variation for a range of
choices of $\Lambda$ as shown in Fig.\ \ref{ScalingF2toF1}. Since we do not
expect the asymptotic predictions for $F_{1,2}(Q^2)$ to work at these $Q^2$,
{\sl the observed consistency might be a sign of precocious scaling as a
consequence of delicate cancellations in the ratio.} A more detailed discussion
on this issue along with more thorough phenomenological analyses will be given
in a separate publication.

We thank V.M. Braun, H. Gao, G.P. Korchemsky, H.-N. Li, and A.V. Radyushkin for
useful discussions. This work was supported by the U. S. Department of Energy via
grant DE-FG02-93ER-40762.


\end{document}